\documentclass[prl,letterpaper,twocolumn,showpacs,superscriptaddress,amsmath,amsfonts,amssymb,preprintnumbers]{revtex4}
\pdfoutput=1
\usepackage[T1]{fontenc}\usepackage[latin1]{inputenc}
\usepackage{dcolumn,graphicx,color,hvfloat,booktabs,microtype,afterpage} \graphicspath{{Figures/}}
\usepackage[charter]{mathdesign}
\renewcommand{\figurename}{Fig.}
\renewcommand{\tablename}{Table}
\makeatletter\renewcommand{\fnum@figure}[1]{\figurename~\thefigure~(color online).}\makeatother
\makeatletter\renewcommand{\fnum@table}[1]{\tablename~\thetable.}\makeatother

\usepackage[colorlinks,plainpages=false,linkcolor=black,urlcolor=blue,citecolor=black,pdfpagemode=UseNone,pdfstartview=FitBH]{hyperref}

\begin{document} \pagestyle{plain}

\title{Weak superconducting pairing and a single isotropic energy gap in stoichiometric LiFeAs}

\author{D.\,S.\,Inosov}\email[Corresponding author: \vspace{4pt}]{d.inosov@fkf.mpg.de}
\affiliation{Max Planck Institute for Solid State Research, Heisenbergstra{\ss}e~1, D-70569 Stuttgart, Germany}

\author{J.\,S.\,White}
\affiliation{Laboratory for Neutron Scattering, Paul Scherrer Institut, 5232 Villigen PSI, Switzerland}

\author{D.\,V.~Evtushinsky}
\affiliation{Leibnitz Institute for Solid State Research, IFW Dresden, D-01171 Dresden, Germany}

\author{I.\,V.~Morozov}
\affiliation{Moscow State University, Moscow 119991, Russia}
\affiliation{Leibnitz Institute for Solid State Research, IFW Dresden, D-01171 Dresden, Germany}

\author{A.\,Cameron}
\affiliation{School of Physics and Astronomy, University of Birmingham, Edgbaston, Birmingham, B15~2TT, UK}

\author{U.\,Stockert}
\affiliation{Leibnitz Institute for Solid State Research, IFW Dresden, D-01171 Dresden, Germany}

\author{V.~B.\,Zabolotnyy}
\affiliation{Leibnitz Institute for Solid State Research, IFW Dresden, D-01171 Dresden, Germany}

\author{T.~K.~Kim}
\affiliation{Leibnitz Institute for Solid State Research, IFW Dresden, D-01171 Dresden, Germany}

\author{A.\,A.\,Kordyuk}
\affiliation{Leibnitz Institute for Solid State Research, IFW Dresden, D-01171 Dresden, Germany}
\affiliation{Institute for Metal Physics of the National Academy of Sciences of Ukraine, 03142 Kyiv, Ukraine}

\author{S.\,V.~Borisenko}
\affiliation{Leibnitz Institute for Solid State Research, IFW Dresden, D-01171 Dresden, Germany}

\author{E.\,M.\,Forgan}
\affiliation{School of Physics and Astronomy, University of Birmingham, Edgbaston, Birmingham, B15~2TT, UK}

\author{R.\,Klingeler}
\affiliation{Leibnitz Institute for Solid State Research, IFW Dresden, D-01171 Dresden, Germany}

\author{J.\,T.~Park}
\affiliation{Max Planck Institute for Solid State Research, Heisenbergstra{\ss}e~1, D-70569 Stuttgart, Germany}

\author{S.~Wurmehl}
\affiliation{Leibnitz Institute for Solid State Research, IFW Dresden, D-01171 Dresden, Germany}

\author{A.\,N.\,Vasiliev}
\affiliation{Moscow State University, Moscow 119991, Russia}

\author{G.\,Behr}
\affiliation{Leibnitz Institute for Solid State Research, IFW Dresden, D-01171 Dresden, Germany}

\author{C.\,D.\,Dewhurst}
\affiliation{Institut Laue-Langevin, 6 Rue Jules Horowitz, F-38042 Grenoble, France}

\author{V.~Hinkov}
\affiliation{Max Planck Institute for Solid State Research, Heisenbergstra{\ss}e~1, D-70569 Stuttgart, Germany}




\begin{abstract}

\noindent We report superconducting (SC) properties of stoichiometric LiFeAs ($T_{\rm c}=17$\,K) studied by small-angle neutron scattering (SANS) and angle-resolved photoemission (ARPES). Although the vortex lattice exhibits no long-range order, well-defined SANS rocking curves indicate better ordering than in chemically doped 122-compounds. The London penetration depth $\lambda_{a\kern-.5pt b}(0)=210\pm20$\,nm, determined from the magnetic field dependence of the form factor, is compared to that calculated from the ARPES band structure with no adjustable parameters. The temperature dependence of $\lambda_{a\kern-.5pt b}$ is best described by a single isotropic SC gap $\Delta_0=3.0\pm0.2$\,meV, which agrees with the ARPES value of $\Delta^{\rm ARPES}_0=3.1\pm0.3$\,meV and corresponds to the ratio $2\Delta/k_{\rm B}T_{\rm c}=4.1\pm0.3$, approaching the weak-coupling limit predicted by the BCS theory. This classifies LiFeAs as a weakly coupled single-gap superconductor, similar to conventional metals.

\end{abstract}

\maketitle\enlargethispage{0.3em}

\noindent In many of the recently discovered Fe-based superconductors (SC) \cite{KamiharaWatanabe08, RotterTegel08}, a transition to the SC state is induced by chemical doping of a parent compound that at ambient conditions does not exhibit SC in its stoichiometric composition even at the lowest temperatures. Among the few known exceptions, the present record holder for the SC transition temperature, $T_{\rm c}$, is the stoichiometric LiFeAs ($T_{\rm c}\kern-1pt\lesssim\kern-1pt18$\,K) \cite{WangLiu08, PitcherParker08, TappTang08}. Others are low-$T_{\rm c}$ superconductors NaFeAs ($T_{\rm c}=9$\,K) \cite{ParkerPitcher09}, FeSe ($T_{\rm c}=9$\,K) \cite{ImaiAhilan09}, LaFePO ($T_{\rm c}=6$\,K) \cite{KamiharaHiramatsu06, FletcherSerafin09}, and KFe$_2$As$_2$ ($T_{\rm c}=3.8$\,K) \cite{SasmalLv08}. The electronic structure of LiFeAs is quasi two-dimensional (2D) \cite{BandStructure} and supports superconductivity in the absence of any notable Fermi surface (FS) nesting or static magnetism \cite{BorisenkoZabolotnyy09}. However, the presence of normal-state antiferromagnetic fluctuations has been suggested by $^{75}\!$As NMR measurements \cite{JeglicPotocnik10}. Together with the weakness of the electron-phonon coupling predicted by the density functional theory \cite{JishiAlyahyaei10}, this suggests that the SC pairing in this structurally simple compound possibly has the same magnetic origin as in higher-$T_{\rm c}$ iron pnictides \cite{MazinSingh08, KurokiOnari08, InosovPark10}. On the other hand, arguments advocating the phonon mechanism have also been raised recently \cite{KordyukZabolotnyy10}. Therefore, to pinpoint the SC mechanism with certainty, details of the SC pairing symmetry and the coupling strength are required.

In a number of recent studies \cite{EskildsenVinnikov09, YinZech09, MagneticPinning, InosovShapoval10}, it was shown that doped iron arsenide superconductors are characterized by strong pinning of magnetic flux lines that precludes the formation of an ordered Abrikosov lattice. The role of the pinning centers can be played by magnetic/structural domains in the underdoped samples \cite{MagneticPinning}, by the dopant atoms themselves, such as Co or Ni, at higher doping levels \cite{InosovShapoval10}, or by the electronic inhomogeneities that result from phase separation in some hole-doped 122-systems \cite{ParkInosov09}. This served as our motivation to study the magnetic field penetration in a single crystal of stoichiometric LiFeAs, which possesses a non-magnetic ground state with tetragonal crystal symmetry, thus excluding all of the above-mentioned strong pinning mechanisms from consideration. In the following, we will compare these results with ARPES measurements of the electronic structure to establish the microscopic origin of the measured quantities.

\begin{figure}[b]\vspace{-1em}
\includegraphics[width=\columnwidth]{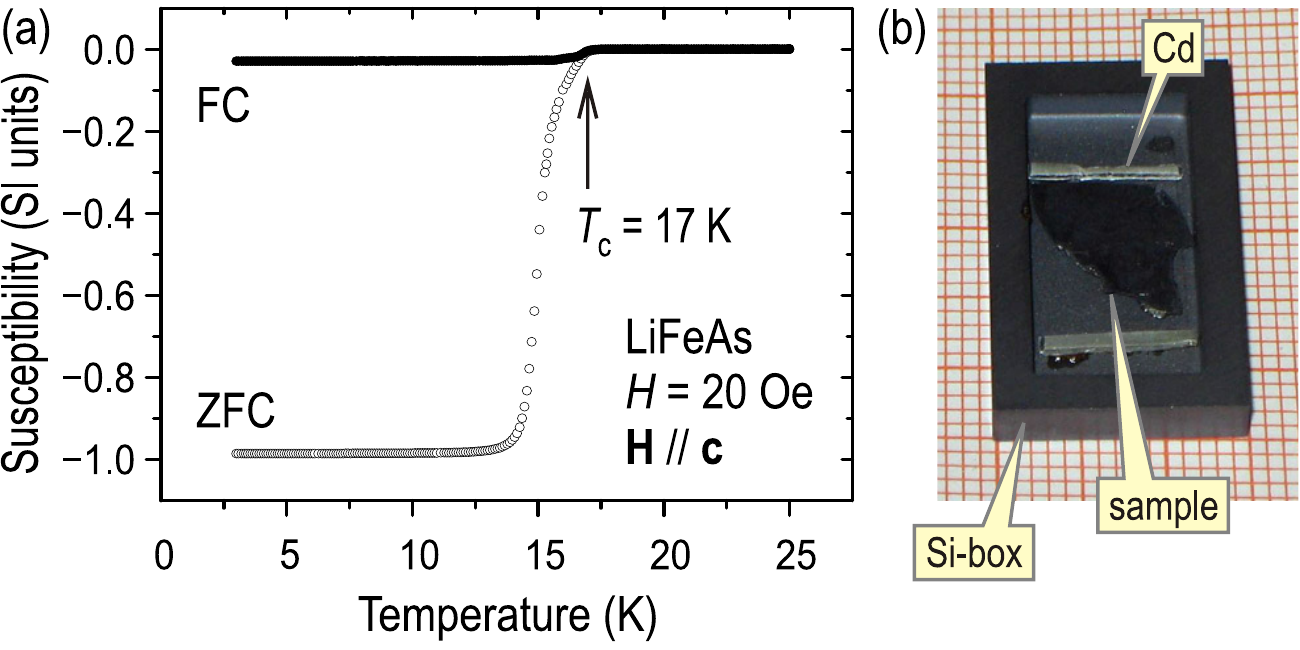}\vspace{-0.5em}
\caption{(a)~Magnetic susceptibility of LiFeAs, measured upon warming after cooling in magnetic field (FC) and in zero field (ZFC). (b)~Photo of the sample prepared for SANS measurements inside the single-crystalline silicon box (see text).\vspace{-1.3em}}
\label{Fig:Magnetization}
\end{figure}

\begin{figure}[t]
\includegraphics[width=\columnwidth]{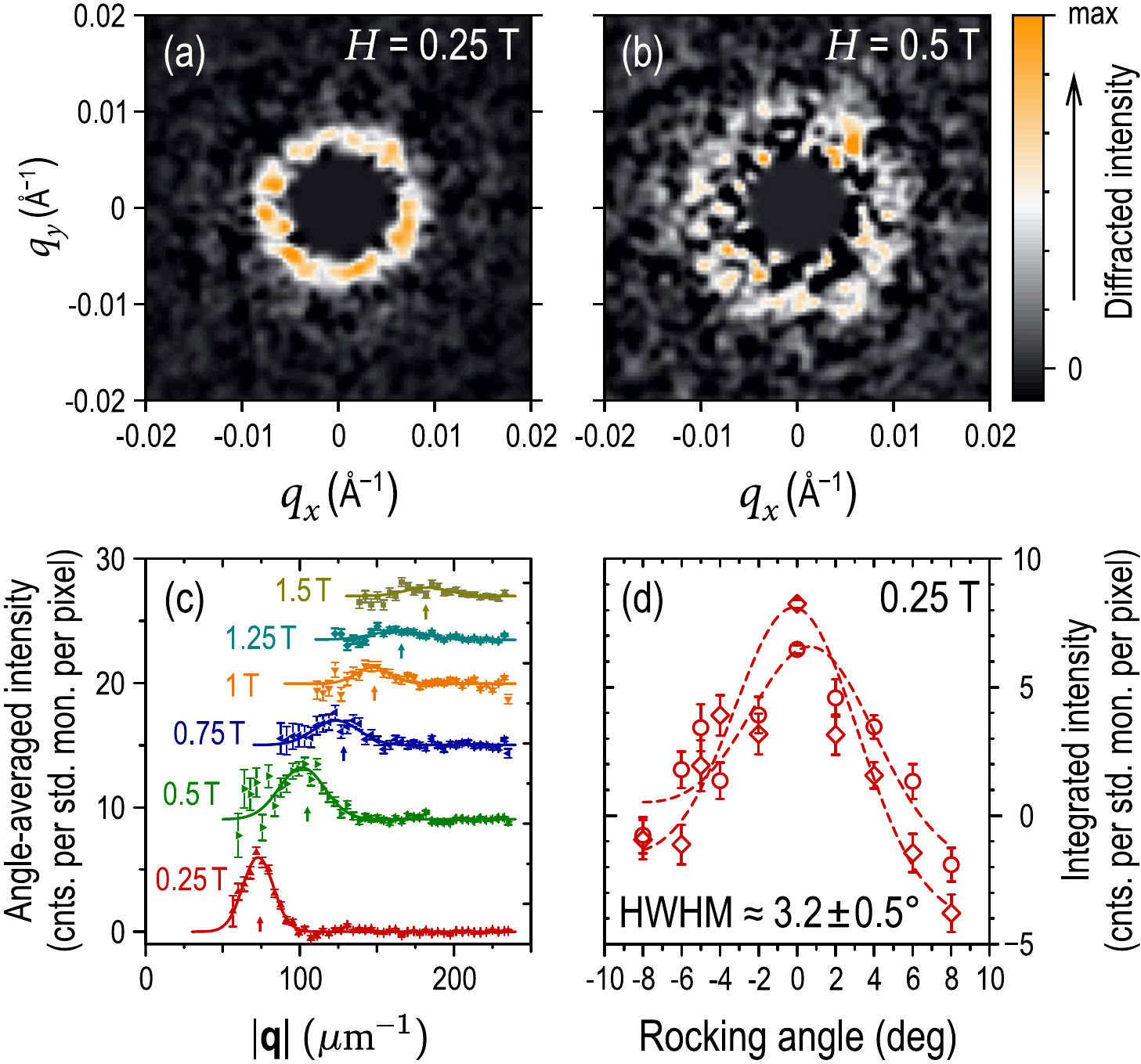}
\caption{(a,\,b) SANS diffraction patterns measured at $H=0.25$ and 0.5\,T, respectively. The 0.25\,T data are summed up over the rocking curve from $-8^\circ$ to $8^\circ\!$, whereas the 0.5\,T data are shown for the zero rocking angle. Both datasets are smoothed with a 3\hspace{-0.1pt}-\hspace{-0.2pt}pixel FWHM Gaussian filter. (c)~Angle-averaged diffracted intensity as a function of momentum transfer $|\mathbf{q}|$, measured at different magnetic fields between 0.25\,T and 1.5\,T. For clarity, the zero line of each curve is offset from the one below it. Vertical arrows show the expected peak positions for a perfect triangular lattice $q_\triangle$. Solid lines are Gaussian fits. (d)~Averaged intensities on the left ($\diamond$) and right ($\circ$) sides of the ring as functions of the rocking angle, measured at $H=0.25$\,T.\vspace{-1.3em}}
\label{Fig:SANS}
\end{figure}

For our SANS experiments, we used a large single crystal with a volume of $\sim\!10\times10\times0.4$\,mm$^3$. It was grown by the self-flux method and characterized as described in Ref.\,\onlinecite{BorisenkoZabolotnyy09}. The value of $T_{\rm c}$, measured on a smaller piece of the same sample, was $\sim\!17$\,K, as shown in Fig.\,\ref{Fig:Magnetization}\,(a). Magnetization data are corrected for demagnetization by an ellipsoid approximation \cite{Osborn45}. In order to avoid any exposure of the sample to air during mounting into the cryostat, it was placed inside a single-crystalline Si box [Fig.\,\ref{Fig:Magnetization}\,(b)]. The box was sealed with a thin rectangular lid prepared from a Si wafer that was glued on top with a small amount of low-temperature varnish inside the glove box with Ar atmosphere. Because of the low absorption and incoherent neutron-scattering cross-sections of Si, such box is essentially transparent to neutrons and has a negligible effect on the background. To mark the position of the sample inside the sealed box with respect to the neutron beam, two stripes of highly neutron-absorbent Cd were placed above and below it.

The SANS experiment was carried out using the D11 instrument at the Institut Laue-Langevin (ILL) in Grenoble. We used the usual experimental geometry, with the applied field approximately parallel to the incident cold-neutron beam, characterized by the wavelength $\lambda=8$\,\AA\ and the full width at half maximum (FWHM) wavelength spread $\Delta\lambda/\lambda=10$\%. The diffracted neutrons were collected by a 2D multidetector placed behind the sample. The vortex lattice (VL) was prepared in the sample by applying the desired field above $T_{\rm c}$, and subsequent field-cooling to 2\,K. In all cases, background measurements were carried out at 20\,K (above $T_{\rm c}$) and subtracted from the field-cooled foreground measurements.

Two representative diffraction patterns measured in magnetic fields $H=0.25$ and 0.5\,T are shown in Fig.\,\ref{Fig:SANS}\,(a,\,b). A distinct ring of scattering is seen in both panels. The absence of distinct Bragg peaks indicates a lack of long-range orientational order of the VL. The 0.25\,T image is a sum of 10 individual diffraction patterns measured at different rocking angles between $-8^\circ$ to $8^\circ\!$, whereas the 0.5\,T image was measured at zero rocking angle only. At higher fields, the scattering intensity is decreased, but a peak can be seen in the angle-averaged data up to $H=1.5$\,T, as shown in panel (c). For all fields, the fitted peak position agrees within the experimental error with the expected radius of the diffraction ring $q_\triangle\!=4\piup/a_\triangle\!\sqrt{3}$, calculated for a perfect triangular VL with lattice parameter \mbox{$a_\triangle\!=\sqrt{2\phi_0/H\sqrt{3}}$, where} $\phi_0$ is the magnetic flux quantum. Calculated positions of $q_\triangle$ are shown in the figure by small vertical arrows.

So far, these results are similar to those reported for electron-doped 122-compounds \cite{EskildsenVinnikov09, InosovShapoval10}. The first essential difference is illustrated in Fig.\,\ref{Fig:SANS}\,(d), which shows rocking curves with half-width at half-maximum (HWHM) of $3.2\pm0.5^\circ\!$, measured at 0.25\,T. They represent the angular dependence of the diffracted intensity on the left and right sides of the diffraction ring. Despite the clear intensity variation, disorder-induced pinning causes the rocking curves to remain broad, and using the HWHM we estimate the longitudinal correlation length of the vortices as $\zeta_\parallel\approx0.8\,\mu$m.

Conversely, in previous SANS experiments on both underdoped \cite{EskildsenVinnikov09} and overdoped \cite{InosovShapoval10} BaFe$_{2-x}$Co$_x$As$_2$, the rocking curves were much broader and extended beyond the measurable range. Our observations therefore indicate an improvement in the longitudinal VL ordering and a decrease of the typical pinning forces in the absence of chemical dopants. But the fact that we did not observe any long-range orientational VL order even after oscillating the field value by 1 and 10\% during in-field cooling suggests that the pinning in our sample is still not negligible.

Now let us turn to the quantitative determination of some important SC properties of LiFeAs. The integrated intensity $I$ corresponding to 1/6 of the diffraction ring (one Bragg spot of a triangular VL), obtained from the rocking curve, is proportional to the modulus squared of the VL form factor $F(q,\!T)$ \cite{ChristenTasset77}, i.e. the Fourier transform of the 2D magnetic-flux modulation within the sample. As our measured intensity mainly originates from the first-order Bragg spots at a distance $q\approx q_\triangle$ from the origin, we limit our considerations to the first order VL form factor,\vspace{-0.2em}
\begin{equation}
I=2\pi V\Phi(\gamma/4)^2\lambda^2\phi_0^{-2}q^{-1}|F(q,\!T)|^2.\vspace{-0.3em}
\end{equation}
Here $V$ is the sample volume, $\Phi$ is the neutron flux density, and $\gamma$ is the magnetic moment of the neutron in nuclear magnetons. By varying the magnetic field and, consequently, $q_\triangle$, we can thus study the $q$-dependence of the form factor, assuming that the rocking curve width is both field- and temperature-independent.

\makeatletter\renewcommand{\fnum@figure}[1]{\figurename~\thefigure.}\makeatother
\begin{figure}[t]
\includegraphics[width=\columnwidth]{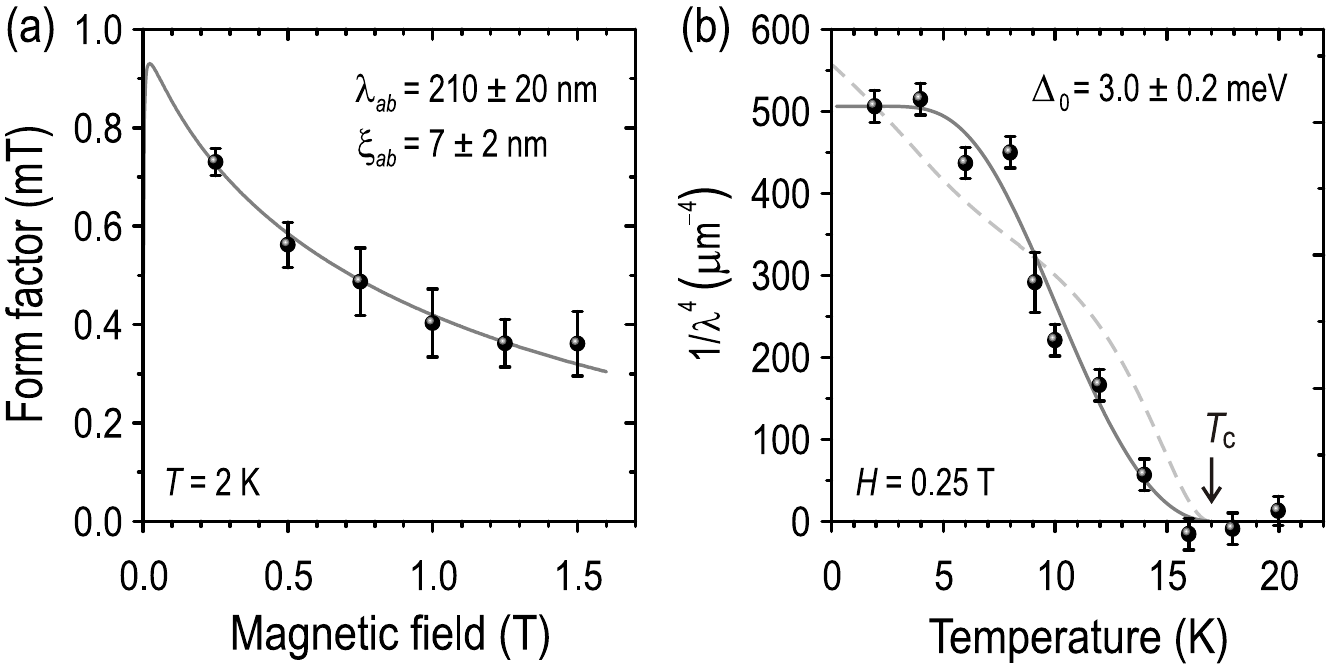}\vspace{-0.5em}
\caption{(a) Magnetic field dependence of the VL form factor at $T=2$\,K, fitted to Eq.\,(\ref{Eq:FormFactor}). (b) Temperature dependence of $|F(q,\!T)|^2$ at $H=0.25$\,T. The vertical axis is scaled to the value of $\lambda^{-4}(0)$ that resulted from the form-factor fit in panel (a). For comparison, $\lambda^{-4}(T)$ for the $d$-wave gap is shown by the dashed line.\vspace{-1.3em}}
\label{Fig:Gap}
\end{figure}

In the simplest Ginzburg-Landau model valid for superconductors with large $\kappa=\lambda/\xi\gg1$, and in small magnetic fields $H\ll H_{\rm c2}$, the form factor can be expressed in terms of the temperature-dependent penetration depth $\lambda_{a\kern-.5pt b}\text{(}T\text{)}$ and the SC coherence length $\xi_{\kern-.5pt a\kern-.5pt b}\text{(}T\text{)}$ \cite{YaouancDalmas97} (for brevity, the index $a\kern-.7pt b$ will be skipped),\vspace{-0.3em}
\begin{equation}
\hspace{-0.5em}F(q,\!T)=\frac{HgK_1\!\text{(}g\text{)}}{1+[\lambda\text{(}T\text{)}\kern.5pt q]^2},~g\!=\!\sqrt{2}~\frac{\xi\text{(}T\text{)}}{\lambda\text{(}T\text{)}}\sqrt{1+[\lambda\text{(}T\text{)}\kern.5pt q]^2}.\!\vspace{-0.3em}
\label{Eq:FormFactor}
\end{equation}
Here $K_1\!\text{(}g\text{)}$ is the modified Bessel function of the second kind
\footnote{Various commonly used approximations to Eq.\,(\ref{Eq:FormFactor}), such as $gK_1\!\text{(}g\text{)}=\mathrm{e}^{-\sqrt{2}\kern.5pt\xi q}$ or $gK_1\!\text{(}g\text{)}=\mathrm{e}^{-2\kern.5pt\xi^2q^2}$ \cite{YaouancDalmas97} resulted in acceptably accurate fits for $\lambda$ and $\Delta$, but turned out to be inaccurate for estimating $\xi$, sometimes nearly two-fold. We have therefore avoided using these approximations.}. Because the band structure of LiFeAs can potentially support two-gap SC \cite{BorisenkoZabolotnyy09}, we resort to the following analytical two-gap model to describe the temperature dependence of the penetration depth,
\begin{equation}
\hspace{-0.5em}\frac{1}{\lambda^2\text{(}T\text{)}}=I_1\!\left[\!1-M\Biggl(\!\!\frac{\Delta_1\text{(}T\text{)}}{k_{\rm B}T}\!\!\Biggr)\!\right]\!+I_2\!\left[\!1-M\Biggl(\!\!\frac{\Delta_2\text{(}T\text{)}}{k_{\rm B}T}\!\!\Biggr)\!\right],\label{Eq:PenetrationDepth}
\end{equation}
where the function $M$ is an accurate approximation \cite{EvtushinskyInosov09} for the temperature-dependent quasiparticle effects, and the constants $I_{1,\,2}$ depend only on the band structure and can be expressed as simple integrals over the FS \cite{EvtushinskyInosov09, KhasanovEvtushinsky09}. The $T$-dependence of the SC gap is approximated by \cite{GrossChandrasekhar86}\vspace{-0.3em}
\begin{equation}
\Delta(T)=\Delta_0\,\tanh\biggl(\!\frac{\piup}{2}\sqrt{T_{\rm c}/T-1}\,\biggr).\vspace{-0.3em}
\label{Eq:TdepGap}
\end{equation}
Consequently, because in the weak-coupling limit the SC coherence length is inversely proportional to the energy gap \cite{BenfattoToschi02}, the temperature evolution of $\xi$ will be given by\vspace{-0.3em}
\begin{equation}
\xi(T)=\xi(0)\biggl[\tanh\biggl(\!\frac{\piup}{2}\sqrt{T_{\rm c}/T-1}\,\biggr)\biggr]^{-1}\!\!\propto1/\Delta(T).\vspace{-0.3em}
\label{Eq:TdepCohLength}
\end{equation}
Substituting expressions (\ref{Eq:PenetrationDepth}) through (\ref{Eq:TdepCohLength}) into Eq.\,(\ref{Eq:FormFactor}), we obtain the final formula for fitting the SANS data.

Following the methodology of Ref.\,\onlinecite{WhiteForgan08}, we start with the magnetic field dependence of the low-temperature form factor that is shown in Fig.\,\ref{Fig:Gap}\,(a). Because at $T=2$\,K the values of $\xi$ and $\lambda$ can be considered equal to their zero-temperature limits, Eq.\,(\ref{Eq:FormFactor}) can be applied directly to the field-dependent data (solid line in the figure) to extract the values of $\lambda_{a\kern-.5pt b}(0)=210\pm20$\,nm and $\xi_{a\kern-.5pt b}(0)=7\pm2$\,nm ($\kappa=29\pm7$). The obtained value of $\lambda_{a\kern-.5pt b}$ agrees with the results of a muon-spin rotation ($\mu$SR) measurement \cite{PrattBaker09}, which yielded $\lambda_{a\kern-.5pt b}(0)=195$ and 244\,nm for two samples of Li$_{1+\delta}$FeAs with $T_{\rm c}=16$ and 12\,K, respectively. Our value of $\xi_{a\kern-.5pt b}(0)$, however, is likely to be overestimated with respect to that obtained from upper-critical-field measurements \cite{SongGhim10}, which is $\xi_{a\kern-.5pt b}(0)=\sqrt{\phi_0/2\piup H^\perp_{\rm c2}}\approx2\!-\!4$\,nm. Such overestimation can result either from a finite-width $T_{\rm c}$ distribution in our large sample, or from the field-induced disorder of the VL expected in the Bragg glass model \cite{KleinJoumard01}.

\makeatletter\renewcommand{\fnum@figure}[1]{\figurename~\thefigure~(color online).}\makeatother
\begin{figure}[b]\vspace{-1em}
\includegraphics[width=\columnwidth]{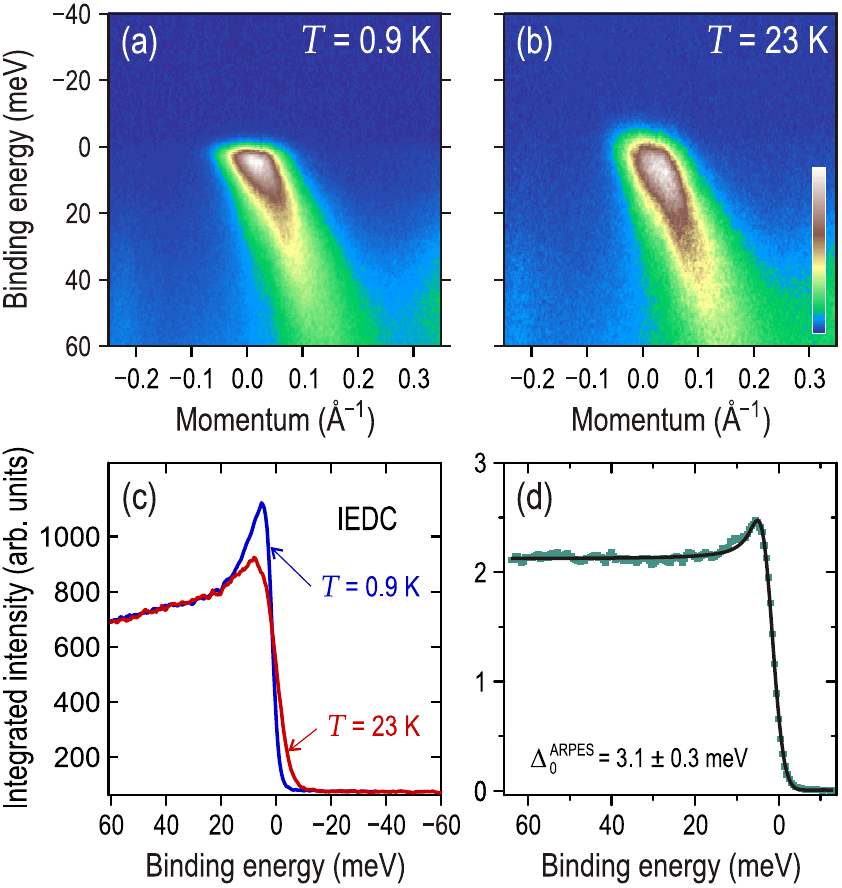}\vspace{-0.5em}
\caption{ARPES spectra of LiFeAs measured on the double-walled electron-like M-barrel in the SC (a) and normal (b) states. (c) The integrated energy distribution curves (IEDCs) of the same spectra. (d)~The low-temperature IEDC after normalization, fitted to the Dynes function.\vspace{-1.3em}}
\label{Fig:ARPES}
\end{figure}

Next, we turn to the temperature evolution of the form factor. As follows from Eq.\,(\ref{Eq:FormFactor}), for $\lambda(T)\kern.5pt q\gg1$, $F(q,\!T)\propto\lambda^{-2}(T)$. The scattered intensity therefore scales $\propto\!\lambda^{-4}$. In Fig.\,\ref{Fig:Gap}\,(b), the measured integrated intensity is plotted as a function of temperature, and the vertical axis is scaled to the value of $1/\lambda^4(0)$ that resulted from the low-temperature fit of the form factor. By fixing $\lambda(0)$ and $\xi(0)$ to the values found previously, we can now fit the two SC gaps $\Delta_{1,2}(0)$ and the coefficients $I_{1,2}$, using $|F(q,\!T)|^2$ as the fitting function. It turns out that independently of the parameter initialization, the fit converges to a single value of the gap $\Delta_1=\Delta_2=3.0\pm0.2$\,meV. This value of the energy gap corresponds to the ratio $2\Delta/k_{\rm B}T_{\rm c}=4.1\pm0.3$, approaching the weak-coupling limit of 3.53 predicted by the BCS theory of conventional superconductivity \cite{BCS57}. For comparison, $\lambda^{-4}(T)$ corresponding to a $d$-wave gap is also shown in the same figure, producing a poor fit. This essentially excludes the possibility of two-gap SC or gap nodes in LiFeAs.

Now we compare these results with those of ARPES, to establish their relationship to the microscopic electronic properties, such as band dispersion and the SC gap. An analysis of the leading edge shift along the FS contours implies an isotropic gap for every FS sheet \cite{BorisenkoZabolotnyy09}. To quantify the low-temperature gap value $\Delta_0$, we employed the Dynes function fitting procedure \cite{DynesNarayanamurti78} to the ARPES spectra measured on the double-walled electron-like M-barrel [Fig.\,\ref{Fig:ARPES}\,(a,\,b)]. The energy distribution curves integrated in a wide momentum window along the FS radius (IEDCs), measured in the SC state below 1\,K and in the normal state at 23\,K, are shown in Fig.\,\ref{Fig:ARPES}\,(c). In order to reveal the true shape of the spectrum in the SC state, the low-temperature IEDC was normalized by the Fermi-function-corrected normal state spectrum, as shown in Fig.\,\ref{Fig:ARPES}\,(d). The good quality of the Dynes-function fit confirms the robustness of such normalization. The resulting low-temperature value of $\Delta^{\rm ARPES}_0=3.1\pm0.3$\,meV is in perfect agreement with that extracted above from the temperature dependence of $\lambda_{a\kern-.5pt b}$.

The knowledge of the band dispersion together with the SC gap allows the calculation of macroscopic properties in the SC state with no adjustable parameters. The superfluid density at $T\rightarrow0$ is proportional to the integral of Fermi velocity $v_{\rm F}$ along the FS perimeter \cite{ChandrasekharEinzel93, EvtushinskyInosov09, KhasanovEvtushinsky09}, and in the clean limit,\vspace{-0.2em}
\begin{equation}
\frac{1}{\lambda_{a\kern-.5pt b}^2} = \frac{e^2}{2\piup\varepsilon_0c^2hL_c} \oint_{\rm FS}\!v_{\rm F}\,{\rm d}k,
\label{Eq:Lambda}
\end{equation}
where $\varepsilon_0$, $h$, $e$, $c$ are physical constants, and $L_c$ is the $c$-axis lattice parameter. Although the FS of LiFeAs consists of several electron- and hole-like sheets \cite{BorisenkoZabolotnyy09}, for the evaluation of the integral (\ref{Eq:Lambda}) the \emph{renormalized} Fermi velocity, extracted from ARPES data, can be well approximated by its average value of $\hbar\langle v_{\rm F}\rangle = 0.3\pm0.03$\,eV\hspace{.5pt}\AA. For the experimental LiFeAs band structure, this formula yields $\lambda^{\rm ARPES}_{a\kern-.5pt b}=172\pm20$\,nm, which is only slightly lower than our directly measured value. Similarly, at $T\rightarrow0$, the BCS coherence length is proportional to the ratio of Fermi velocity to gap magnitude, $\xi^{\rm ARPES}_{a\kern-.5pt b}(0) = \hbar\langle v_{\rm F}\rangle/\piup\Delta_0$ \cite{BCS57}, which equals $3.2\pm0.4$\,nm in our case. This corresponds to the upper critical field $H^\perp_{\rm c2}=$ $32\pm8$\,T, in agreement with direct measurements \cite{SongGhim10}.

In summary, we have evaluated several important SC parameters of LiFeAs from two complementary experiments. We have demonstrated that its order parameter is isotropic and in contrast to the higher-$T_{\rm c}$ ferropnictides \cite{EvtushinskyInosov09} is characterized by a single SC gap $\Delta_0=3.0\pm0.2$\,meV. This value is close to the BCS limit of $1.76\,k_{\rm B}T_{\rm c}$, which indicates that LiFeAs is a weakly coupled single-gap superconductor, similar to conventional metals.

We thank B.\,Büchner and B.\,Keimer for their helpful suggestions and support, and acknowledge discussions with L.\,Boeri and S.\,A.\,Kuzmichev. Sample growth was supported by the DFG project BE\,1749/12. SANS experiments were done with financial assistance from the EPSRC UK and MaNEP. I.\,V.~M. acknowledges support from the Ministry of Science and Education of Russian Federation under the State contract P-279. ARPES spectra were measured with the ``$1^3$-ARPES'' end station, using synchrotron radiation from the BESSY\,II storage ring in Berlin.

\end{document}